\documentclass[12pt,preprint]{JHEP3}
\usepackage{amsmath,amssymb,graphicx,mathrsfs}
\usepackage[latin1]{inputenc}
\usepackage{bm}

\newcommand{\be}{\begin{equation}}
\newcommand{\ee}{\end{equation}}
\newcommand{\bea}{\begin{eqnarray}}
\newcommand{\eea}{\end{eqnarray}}
\title{ Unitarity constraints on the ratio of shear viscosity to entropy density in higher derivative gravity }
\author{ Ram Brustein ${}^{(1,2)}$,  A.J.M. Medved ${}^{(3,4)}$  \\
(1)  Department of Physics, Ben-Gurion University,
     Beer-Sheva 84105, Israel,   \\
(2)  CERN, Theory Division, CH-1211,
     Geneva 23, Switzerland \\
(3)  Department of Physics $\&$ Electronics, Rhodes University,
     Grahamstown 6140\\ $\;\;\;\;\;$ South Africa \\
(4)  School of Physics, KIAS,
     Dongdaemun-gu, Seoul 130-722,  Korea \\
   {\small E-mail: ramyb@bgu.ac.il,\ j.medved@ru.ac.za} }
\date{}
\preprint{}
\abstract{

We discuss corrections to the ratio of shear viscosity to entropy density $\eta/s$ in higher-derivative gravity theories. Generically, these theories contain ghost modes with Planck-scale masses. Motivated by general considerations about unitarity, we propose new boundary conditions for the equations of motion of the graviton perturbations that force the amplitude of the ghosts modes to vanish.
We analyze explicitly four-derivative perturbative corrections to Einstein gravity which generically lead to four-derivative equations of motion, compare our choice of  boundary conditions to previous proposals and show that, with our  new prescription, the ratio $\eta/s$ remains at the Einstein-gravity value of $1/4\pi$ to leading order in the corrections. It is argued that, when the new boundary conditions are imposed on six  and higher-derivative equations of motion, $\eta/s$ can only increase from the Einstein-gravity value. We also recall some general arguments that support the validity of our results to all orders in the strength of the corrections to Einstein gravity. We then discuss the particular case of Gauss-Bonnet gravity, for  which the equations of motion are only of two-derivative order and the value of $\eta/s$ can decrease below $1/4\pi$ when treated in a nonperturbative way. Our findings provide further evidence for the validity of the KSS bound for theories that can be viewed as perturbative corrections to Einstein Gravity.
}

\begin{document}

\section{Introduction}
\label{intro}

The gauge--gravity duality \cite{Maldacena,Witten} allows one to
holographically  map black brane thermodynamics and hydrodynamics in the Anti-deSitter (AdS) bulk to their gauge-theory correspondents at the AdS boundary \cite{KSS-hep-th/0309213,SS-0704.0240}.

For a large class of strongly coupled  fluids  (essentially, any with a two-derivative gravity dual), the ratio of the shear viscosity $\eta$ to the entropy density $s$ is, in appropriately chosen units, remarkably low:
$\;\frac{\eta}{s}=\frac{1}{4\pi}\;$
\cite{PSS-hep-th/0104066,KSS-hep-th/0405231}.
The uncertainty principle can be used to argue  that $\eta/s$ should have a lower bound of order unity  \cite{KSS-hep-th/0405231}; leading
Kovtun, Son and Starinets (KSS) to propose that  $1/4\pi$ is a universal lower bound on this ratio. However, investigations of higher-derivative gravity theories have revealed that the so-called KSS bound can apparently be violated \cite{BLMSY-0712.0805,same-day}. Many subsequent inquiries have appealed to gauge-theory causality as the physical principle that bounds
$\eta/s$  ({\it e.g.}) \cite{BLMSY-sequel}-\cite{Myers09}.
%Hoffman1,Hoffman2

We  have  recently asserted  \cite{us-new} that a unitary ghost-free extension
of Einstein gravity can, at best, saturate  the KSS bound. We have observed that $\eta$ is a gravitational coupling and any such coupling can only increase from its Einstein value in unitary theories \cite{Dvali1,Dvali2,Dvali3}. The source of tension between our claim and the apparently bound-violating gravity theories is that, even though  higher-derivative gravity theories contain ghosts, these are typically at the Planck scale and  therefore expected to be irrelevant to the calculation of hydrodynamic transport coefficients like $\eta$.

Let us briefly recall why ghosts are problematic for any classical or
quantum field theory (also see, {\it e.g.}, \cite{ghost1}-\cite{ghost5}).
If ghosts are present, then any state, including the vacuum,
would be catastrophically unstable due to the  spontaneous creation
of positive and negative energy particles having zero total energy.
For a classical theory of gravity, in particular,
ghost gravitons lead to instabilities at arbitrarily short time scales or
superluminal propagation in both the gravitational and matter sectors.
These issues have to do with the wrong sign of the kinetic energy and not with  the
relative sign of the kinetic and mass terms. Although
the latter can, if ``incorrectly'' chosen, have further effects that  undermine both stability and causality.

The necessity to remove ghosts becomes even  more acute
when the gauge--gravity duality is considered.
From the field-theory side, only a finite number
of states is permissible, which corresponds to a truncated  spectrum
of  bulk excitations. This so-called ``stringy exclusion principle''
\cite{ghost6} has been argued to prohibit ghosts \cite{ghost7}.

From an effective-theory perspective,
the situation is  not so clear cut when all the ghosts  have a  mass on the
order of the ultraviolet cutoff. One might then  reason that it is
now safe to disregard their presence. This could be true in many instances,
but this reasoning will be  shown to be incorrect for a calculation
of $\eta$.  So, to proceed in a sensible manner,  one should  ensure that the ghost
has decoupled before the computation is carried out. This could be accomplished through a choice of boundary conditions (BC's).

Later on, we use four-derivative theories of gravity to explicitly show that the Planck-scale ghosts do indeed infiltrate the existing schemes of calculating $\eta$. Then, motivated by considerations of unitarity,  we propose a new set of BC's for  the higher-order equations of motion that forces the amplitude of the ghost modes to vanish. Applying the new prescription to four-derivative gravity
theories, we find that $\eta/s=1/4\pi$ to leading order in the strength of the corrections to Einstein gravity. We then argue that, after imposing the new BC's on  six and higher-derivative equations of motion,  $\eta/s$ can only increase above $1/4\pi$. These results support our previous claim that $1/4\pi$ is indeed a lower bound on $\eta/s$ for any unitary weakly coupled extension of Einstein  gravity.

Let us now suppose that  a gauge field theory is given and its AdS gravity dual is known. So,  just how does one go about determining $\eta/s$ from a black brane theory on the  AdS side?   For any two-derivative theory of gravity,~\footnote{The two-derivative  class is  meant to include theories that are effectively
two-derivative; {\it e.g.}, any $f({\cal R})$  theory can be expressed as
Einstein gravity plus a scalar.}
this is a well understood matter, and the resulting answer of any method must agree with the well-known
Einstein result of $\;\eta/s=1/4\pi\;$
\cite{PSS-hep-th/0104066,KSS-hep-th/0405231}.
But, for higher-derivative gravity theories, the situation
is still quite ambiguous.

The Wald Noether-charge formalism \cite{wald1,wald2} is appropriate and accepted for the calculation of the entropy or its density $s$.  For the shear viscosity $\eta$,  the situation is less clear. The earliest calculations
of  $\eta$ for Einstein gravity used holographic techniques
to express the field-theory viscosity --- which follows from the Kubo formula ---
in terms of horizon-valued parameters of the AdS brane theory. (For a review
and references, see \cite{SS-0704.0240}.)
Later, the physical connection between
the boundary  and black brane theories
has  been made more explicit via the membrane-fluid interpretation of horizon hydrodynamics \cite{KSS-hep-th/0309213}.

For any two-derivative gravity theory, one can also obtain $\eta$ from  the horizon value of the coefficient  of the kinetic term of the transversely polarized gravitons \cite{BGH-0712.3206}-\cite{CAI}. This is the same as reading off the coefficient   of the graviton propagator  $\langle h_{xy}h_{xy}\rangle$,  where $h_{\mu\nu}$ is the first-order correction to the background metric~\footnote{$x$ and $y$   represent transverse brane directions that are mutually orthogonal to one another, as well as to the propagating direction of gravitons moving along the brane.}.
%BM-0808.3498,LI

For higher-derivative theories, it is reasonable to apply a similar procedure. Except that, in this case, one first requires a well-defined prescription for dealing with  the higher (than second) order derivatives which inevitably turn up in the action. A popular approach has appeared in different yet equivalent guises in the literature \cite{ALEX}-\cite{CLEV}. For instance, according to \cite{DUDE}, one is instructed  to iteratively apply the field equation to the action,
consistently trading off the perturbative-order higher derivatives for lower ones  (while ignoring terms of second or higher perturbative order), until
a two-derivative action has been obtained. Then, the viscosity is extracted from the kinetic term of this  effective  action.
Identical results have been obtained from a canonical-momentum formulation of $\eta$ \cite{ROB}, which can be viewed as a higher-derivative generalization of the membrane paradigm  \cite{KSS-hep-th/0309213} as interpreted in  \cite{LI}.
%DUDE,BMS,ROB,

An alternative approach   \cite{BM-0808.3498} prescribes ignoring the higher-order derivatives and extracting $\eta$ straight from the kinetic term. In this approach, $\eta$ is presumed to maintain its identity as the coefficient of the propagator. For theories limited to four derivatives, the two approaches  are in agreement. This agreement does not, however, persist for theories with six or more derivatives,  nor could it be expected to. For the sake of completeness, we will elaborate further  on these methods in
%Section~\ref{section:eta}.
an appendix.
Our goal in this paper is not to argue for the merit of one prescription versus the other. Rather, it is our contention that both of these approaches should be corrected.

The rest of the paper proceeds as follows: In  the next section,
%~\ref{section:eom},
we introduce the explicit  models
which are  four-derivative extensions of Einstein gravity. Here, our conventions are set and much of  the necessary formalism is presented,
with particular emphasis on  deriving the linearized field equations.
In Section~\ref{section:bc}, we present a detailed discussion
on the calculations of the shear viscosity for such higher-derivative theories.
%In Subsection~\ref{section:eta},
A thorough inspection of  the field equations confirms that
 these existing methods are equivalent
to  reading off the shear viscosity from the coefficient of the kinetic term.
Using this observation, we are then able to demonstrate explicitly that ghosts have  infiltrated
the earlier attempts at evaluating $\eta$.
Section~\ref{section:neweta} begins with  a  proposal for new BC's, as part of a revised  prescription for evaluating the shear viscosity.
This choice forces the decoupling of the ghosts that are inherent  to
higher-derivative theories. Subsequently, we explain how our new prescription protects the KSS bound and elaborate on the physical basis for the bound's validity.
In Section~\ref{GBR}, we resolve an apparent contradiction between our findings and the special case of Gauss--Bonnet gravity.
Section~\ref{conclusions}
contains a brief summary of the results and their significance. The paper concludes
with an appendix,
where we
provide a detailed description of  previous methods for
calculating $\eta$.
%and (B) comment on  a  particularly subtle
%aspect of Gauss--Bonnet gravity.

\section{Equations of motion}
\label{section:eom}

\subsection{Specific 4-derivative models}

For clarity and concreteness, we consider the case of
a  $5$-dimensional AdS  black
brane and  focus on  the class of
four-derivative theories of pure gravity.
Let us now set the notation and conventions.
The background metric can be expressed as
\be
ds^2\;=\;-F(r)dt^2+ \frac{dr^2}{F(r)}+\frac{r^2}{L^2}\left[dx^2+dy^2+dz^2\right]\;,
\label{metric}
\ee
where
$\;F(r)=\frac{r^2}{L^2}\left[1-\frac{r_h^4}{r^4}\right]\;$,  while $L$ and $r_h$  respectively denote the radius of curvature and  black brane horizon. The function $F$ vanishes on the horizon, $\;F(r_h)=0\;$.

The theories of interest are completely described by
the Lagrangian  of Einstein gravity
in AdS space,
\begin{equation}
\label{einsteinlag}
\;{\cal L}_E={\cal R}+\frac{12}{L^2}\;,
\end{equation}
and  three perturbative corrections as follows:~\footnote{For
a four-derivative theory, any derivative of the Riemann tensor
reduces to a surface term and can therefore be ignored.}
\begin{align}
\label{Rsquarelag}
\;{\cal L}_A &= \alpha L^2 {\cal R}^2\;, \\
\label{Riccisquarelag}
\;{\cal L}_B & = \beta L^2 {\cal R}_{ab}{\cal R}^{ab}\;, \\
\label{Riemannsquarelag}
\;{\cal L}_C & = \gamma L^2 {\cal R}_{abcd}{\cal R}^{abcd}\;.
\end{align}
The validity  of the gauge--gravity duality mandates the hierarchy
$\;r_h \gg L \gg l_p\;$ (where $l_p$ is the Planck length), thus making
the dimensionless coefficients $\alpha$, $\beta$, $\gamma \sim l_p^2/L^2$
much smaller than unity.
We also consider, with particular emphasis in Section~\ref{GBR}, the special Gauss--Bonnet combination of the three perturbations or
$\;{\cal L}_{GB}\equiv\lambda\left[{\cal L}_A-4{\cal L}_B+{\cal L}_C\right]\;$.

Our  interest is in the case of weak gravity,
and so the linear expansion of the metric about
its background $\;g_{ab}={\overline g}_{ab} +h_{ab} +{\cal O}[h^2]\;$ is valid.  Similarly,
we need linearized expressions for  the various curvature tensors; for instance,
\be
{\cal R}_{abcd}(h) \;=\; \frac{1}{2} \left[{\overline\nabla}_c{\overline\nabla}_b h_{ad}
\;+\;  {\overline\nabla}_d{\overline\nabla}_a h_{bc}
\;-\; {\overline\nabla}_d{\overline\nabla}_b h_{ac}
\;-\; {\overline\nabla}_c{\overline\nabla}_a h_{bd} \right]\;,
\label{R-of-h}
\ee
from which  the contracted forms follow. An overline
indicates the background geometry.

To obtain the  linearized graviton field equations, we use
the following  identity  \cite{wald2,Padmanabhan:2009vy}:
\be
\frac{\delta{\cal L}}{\delta g^{pq}}\;-\;
2\nabla_a\nabla_b\left[\frac{\delta{\cal L}}{\delta{\cal R}_{a\;\;\;\;b}^{\;\;pq\;\;}}\right]
\;+\;{\cal R}_{abcp}\left[\frac{\delta{\cal L}}{\delta{\cal R}_{abc}^{\;\;\;\;\;q}}\right]
\;-\; \frac{1}{2}g_{pq}{\cal L}
\;=\; 0\;.
\label{wald-eq}
\ee
One can obtain the desired equations by expanding out each term to
linear order in the $h$'s. (In our case, the first term is trivially vanishing.)
For the current analysis, this treatment is preferred over the more
familiar procedure
of varying with respect to the metric,  as the absence of explicit derivatives in
the Lagrangian means that  boundary terms need not be stipulated.
To determine $\eta$, we compute the $xy$ component of  Eq.~(\ref{wald-eq}), knowing that the $\{x$,$y\}$ sector does not mix with other
polarizations.  The calculations are carried out  in the transverse, traceless gauge   and, often, at the  horizon.  The brane horizon is the  most appropriate surface  for an analysis of $\eta$ from the gravity  side; however, none
of our  conclusions are sensitive to this choice.
Deriving the  Einstein equation for ${\cal L}_E$ and multiplying through by a factor of $-2$, one then obtains
\be
\Box_E \; h_{xy} \;\equiv\;
\left[\Box_L  +{\overline{\cal R}}+\frac{12}{L^2} \right]h_{xy}\; =\;0\;,
\label{Efeg}
\ee
where
$\;\Box_L \equiv \Box-{\overline {\cal R}}^x_{\ x}-{\overline {\cal R}}^y_{\ y}
-2{\overline {\cal R}}^{xy}_{\;\;\;xy}\;$ and
$\;\Box = {\overline g}^{ab}{\overline\nabla}_a{\overline\nabla}_b\;$. The operator $\Box_L$ is
the spin-2 Lichnerowicz d'Alembertian \cite{Lich}  ({\it i.e.}, the analogue of $\Box$ for a graviton in curved space) for the background metric.
When the field equation is re-expressed in terms of $\phi = {\overline g}^{xx} h_{xy}$, it is  formally equivalent
to the Klein--Gordon equation for a massless scalar
throughout the spacetime.  On the horizon, in particular, Eq.~(\ref{Efeg}) becomes
\be
\;\Box_E \; h_{xy} \;=\;\Box \; h_{xy} \;=\; 0\;.
\label{Efe}
\ee
This observation proves to be important later on.

Let us next consider the leading-order contributions from the corrections.
After tedious but straightforward calculations, we obtain
\begin{align}
\label{Afex}
{\cal L}_A \;&\to\; -40\alpha\Box\; h_{xy}+\cdots\;, \\
\label{Bfeg}
{\cal L}_B\;&\to\; \beta L^2\left[\Box^2 \;-\;
%\left(
2\frac{\partial_r F}{r}
%+ 10\frac{F}{r^2}\right)
\Box\right]  h_{xy} +\cdots \;, \\
{\cal L}_C\;&\to\;  4\gamma L^2\left[\Box^2 \;+\;\left(
-\frac{1}{2}\partial_r^2 F
-\frac{1}{2}\frac{\partial_r F}{r}
+ 7\frac{F}{r^2}\right)\Box\right]  h_{xy}+\cdots \;.
\label{Cfeg}
\end{align}
The ellipsis represent  perturbative-order mass terms, which will
be omitted for the remainder of the paper, as these are inconsequential
to the current discussion. In any event,
one can eliminate these by  expressing the on-shell form of
the full field equation
in terms of $\;\phi=h^x_{\ y}\;$.

The $\Box^2$ terms in the equation of motion are focal to  the ensuing discussion, so let us be more explicit how they arise.  After linearizing  Eq.~(\ref{wald-eq}),
one finds that a $\Box^2$ term  can only come from
$\;{\overline \nabla}_a{\overline \nabla}_b {\cal X}(h)^{a\;\;\;\;\;\;b}_{\;\;\;xy\;\;\;}\;$,
where $\;{\cal X}^{abcd}\equiv \delta{\cal L}/ \delta{\cal R}_{abcd}\;$. Considering, for instance,  the $4$-index Riemann-squared case, one then has
$\;{\cal X}_{\;\;xy}^{a\;\;\;\;b}=2\gamma L^2 R_{\;\;xy\;}^{a\;\;\;\;b}\;$, which
can be linearized via Eq.~(\ref{R-of-h}).
In this case,  only the second term in Eq.~(\ref{R-of-h}) is relevant,
leading to
$\;2\gamma L^2{\overline \nabla}^a\Box{\overline \nabla}_a h_{xy}\;$.
Next,   commutator relations such as
$\;[{\overline \nabla}_a{\rm ,}{\overline \nabla}_b]V_c={\overline {\cal R}}_{abc}^{\;\;\;\;\;d}V_d
\;$
can be utilized to attain (up to mass terms)
$\;2\gamma L^2\left[\Box^2 +{\overline R}_{a}^{\;\;bac}{\overline\nabla}_b{\overline\nabla}_c
\right] h_{xy}\;$ or $\;2\gamma L^2 \left[\Box^2 -\frac{4}{L^2}\Box\right]h_{xy}\;$,
with the latter form following from symmetries of the background as explained below.
Finally, restoring all numerical factors,  one  recovers the $\Box^2$ term in
Eq.~(\ref{Cfeg}).

An important caveat about  ${\cal L}_C$
is that, for this case only, one finds terms
that cannot be expressed directly in terms of $\Box$; with
these  always being of the form
 $\;{\overline {\cal R}}^{axb}_{\;\;\;\;\;x}
{\overline \nabla}_a {\overline \nabla}_b h_{xy}\;$
(as well as the obvious $x\leftrightarrow y$ analogue).
However, we are interested in the $t$ and $r$ sectors because time derivatives are responsible for the  ghosts (if any) and   only  radial derivatives will survive once the hydrodynamic limit is  imposed to calculate $\eta$. Then, since  only  the terms with $\;a,b=\{r,t\}\;$  are relevant
and
$\;{\overline {\cal R}}^{tx}_{\;\;\;\;tx}=
{\overline {\cal R}}^{rx}_{\;\;\;\;rx}=-\frac{1}{2}\frac{\partial_r F}{r}\;$,
  we can   make the  substitution
$\;{\overline {\cal R}}^{axb}_{\;\;\;\;\;x}
{\overline \nabla}_a {\overline \nabla}_b h_{xy}\;
\to\;-\frac{1}{2}\frac{\partial_r F}{r} \Box h_{xy}\;$.
%{\overline {\cal R}}^{rxr}_{\;\;\;\;\;x}
The  remaining  discrepancy
$\;
\left[{\overline {\cal R}}^{zxz}_{\;\;\;\;\;x}
-\left(-\frac{1}{2}\frac{\partial_r F}{r}
%{\overline {\cal R}}^{rxr}_{\;\;\;\;\;x}
\right)\right]
{\overline  \nabla}^z {\overline \nabla}_z  h_{xy}
= \frac{1}{2}\frac{L^2}{r^4}\left[2F-r\partial_r F\right]
k^2 h_{xy}
\;$
(with  $\;k\equiv -i{\overline \nabla_z}\;$) may  be regarded as a
perturbative-order mass term.

The Gauss--Bonnet combination leads to the  correction
\be
{\cal L}_{GB}
\;\to\; -2\lambda L^2 \frac {\partial_r F}{r}\Box\; h_{xy}\;,
%+2\lambda L^2\left[\left(
%\frac{\partial_r F}{r}
%- 4\frac{F}{r^2}\right)\Box\right]  h_{xy}  \;,
\label{GBeg}
\ee
where the background  relations $\;\frac{\partial_r F}{r}+2\frac{F}{r^2} =\frac
{4}{L^2}\;$ and
$\; \partial_r^2 F= 4\frac{F}{r^2}-\frac{\partial_r F}{r}\;$
have been used
to re-express the contributions from ${\cal L}_A$ and ${\cal L}_C$ respectively.
As a useful check, we find that this Gauss--Bonnet correction
reduces in the AdS vacuum (\;$r_h=0$\;)
to $\;-4\lambda\Box  h_{xy}\;$,
in agreement with the known result \cite{BouDes}.

For future reference, the horizon forms of Eqs.~(\ref{Afex}-\ref{GBeg})
are as follows:
\begin{align}
\label{Afe}
{\cal L}_A \;&\to\; -40\alpha\Box\;  h_{xy}\;, \\
\label{Bfe}
{\cal L}_B\;&\to\; \beta L^2\left[\Box^2 \;-\; \frac{8}{L^2}\Box\right]  h_{xy}  \;, \\
\label{Cfe}
{\cal L}_C\;&\to\;  4\gamma L^2
%\left[
\Box^2
%\;+\; \frac{8}{L^2}\Box\right]
\; h_{xy}  \;, \\
{\cal L}_{GB}\;&\to\; -8\lambda\Box\;  h_{xy}\;.
\label{GBfe}
\end{align}

The background metric has been used in the above and any
subsequent computations.
Since our main interest is
in corrections to  $\eta$ up to leading order in $l_p^2/L^2$,
this choice is justified for contributions that come
directly from the perturbative
part of the Lagrangian. But what about
those from the Einstein part? To understand why the background metric still suffices, let us consider the following:
Given the Einstein Lagrangian and some perturbative
correction, the leading-order effect on the near-horizon geometry can  only be to shift
$L$ and $r_h$ from their background values.
We can also, to leading
order,  re-express
any contribution  from the perturbative part
in terms of the same shifted values of $L$ and $r_h$.
Now, if our primary interest is to compute $\eta/s$
for a given theory, then all dependence on $L$ and $r_h$ will  cancel
out of this ratio.  Meanwhile, if our main interest is to compare  $\eta$ for
two different theories, then it is natural to do so at a fixed value of brane temperature
$\;T=r_h/\pi L^2\;$ and, hence, at common values of $L$ and $r_h$.
So that, for current purposes, any correction to the background geometry never does
come into play.

We can restate the above argument in another way.
As already discussed,  one can fully determine $\eta$ for two-derivative
gravity from the coefficient of the kinetic term in the Lagrangian or, equivalently, from the
coefficient of the $\Box$ terms in the field equation.
For the Einstein part of the Lagrangian, the latter coefficient is universally $-1/2$. This is true anywhere in the spacetime and irrespective of the solution. So that, to understand how $\eta$ is corrected for an extended theory,
we need only to determine  how the additional (``non-Einstein'') parts of the Lagrangian
{\em explicitly} contribute. And, with these additions already being of order $l_p^2/L^2$,
the Einstein background metric (\ref{metric}) suffices.

\subsection{Generic 4-derivative gravity}

Let us  emphasize an important property of {\it any}
four-derivative theory of pure gravity
and, in fact, any  higher-derivative extension that
depends only on the Riemann tensor (and its derivatives):
After the common factor of  $L^2$ ({\it cf},
Eqs.~(\ref{Rsquarelag}-\ref{Riemannsquarelag}))  has been factored out of the corrected part of the field equation, the coefficient
of the $\Box^2$  term is a  geometry-independent number   (possibly zero)
and,  hence,  a spacetime invariant. By geometry independent,
we mean that it can depend on the form of the Lagrangian  but
not on the solution to the field equations. This outcome follows
from the observation that  the prefactor for the $\Box^2$ term  is
dimensionless; meaning that it can not depend on the Riemann tensor nor
its derivatives, as any of these  have a  strictly positive mass  dimension (nor can it depend explicitly on the metric, as the Lagrangian is only a function of the curvature). On the other hand, the prefactor for the linear-$\Box$ term has a  dimension of mass squared, and so it can depend on the Riemann tensor.

Consequently,  for a generic four-derivative theory { at any radius $r$},
the corrected field equation can be written  as
\be
\Box_E  h_{xy} \;+\;\epsilon L^2 \left[a\Box^2 +\frac {b(r)}{L^2}\Box\right] h_{xy}
\;=\;0\;,
\label{generic}
\ee
where $\epsilon\sim l_p^2/L^2$ is a perturbative coefficient,  $a$ is a fixed  number of order unity and $b(r)$ is  a radial function of order unity; all of which are dimensionless.  On  a  constant-radius surface, $b$ can also be regarded as a number. We are interested in corrections to $\eta/s$ to leading order in $\epsilon$ and therefore need to consider a finite (small) $\epsilon$.

We need not limit considerations  to the $\{x,y\}$-polarization channel
for the gravitons,
as $\eta$ can also be extracted from the so-called shear and sound channels
(see, {\it e.g.}, Appendix B of \cite{BLMSY-0712.0805}). For the shear channel
(the sound channel follows along similar lines), the relevant gravitons are $h_{ra}$, $h_{ta}$, $h_{za}$ with
$\;a=\left\{x,y\right\}\;$.  The field equations for this class are more complicated  but can be simplified by
choosing the radial gauge $\;h_{r\mu}=0\;$ ($\;\mu=\{r,t,z,a\}\;$) and then considering the gauge-invariant combination
$\;{\cal Z}=g^{xx}k h_{tx}\; + g^{xx} \omega h_{zx}\;$
($\;\omega\equiv +i{\overline \nabla_t}\;$,$\;k\equiv -i{\overline \nabla_z}\;$).
In this way, the coupled set of equations is turned into a single equation for a single graviton mode
${\cal Z}$. This equation is necessarily of the same generic form
as Eq.~(\ref{generic}), as none of the above arguments are specific to the $\{xy\}$-polarization class.
Then, just as we go on to show how the shear viscosity can be extracted from the kinetic coefficient
of  $h_{xy}$, the same can be deduced for the kinetic coefficient of ${\cal Z}$ \cite{Shear} (and analogously for the sound channel \cite{Sound}).

\section{Calculating $\eta$}
\label{section:bc}

\subsection{Contributions to $\eta$}
\label{section:eta}

Next, we  consider the shear viscosity $\eta_X$ for the extended theories ${\cal L}_E +{\cal L}_X\;$, with
$X=\{A$,$B$,$C$,$GB\}$.  But, first, let us recall what is known about
two-derivative theories: The kinetic coefficient of the
transverse gravitons at the horizon fully determines
the shear viscosity for both  the ``graviton fluid'' on the black  brane and its
field-theory dual at the AdS boundary. This follows
from a rigorous calculation in \cite{CAI}
(also, \cite{CAI2})
that applies to
any two-derivative (but otherwise arbitrary) $5D$ gravity theory.
There it was shown that the
Kubo form of the  field-theory viscosity $\eta_{FT}$
can be  expressed purely in terms of  the bulk
geometry. They then imposed
the standard choice   \cite{KSS-hep-th/0309213}  of
incoming plane wave BC  at the horizon and the Dirichlet BC
at the outer boundary to arrive at
\be
\eta_{FT}\;=\;\eta_{grav}\;=\;
\frac{1}{16\pi l_p^3}\frac{r_h^3}{L^3}K(r=r_h)\;,
\label{cai-eqn}
\ee
where $K$ is the  kinetic coefficient of
the transverse gravitons
(here, normalized so that its Einstein value is unity).

Supposedly, we can determine how  each extension modifies the
Einstein viscosity $\eta_E$
by inspecting the horizon forms  of the
corrected
field equations
(\ref{Afe}-\ref{GBfe}).  The kinetic coefficients, in particular, should
tell us how the viscosity is modified.  The scalar-squared
and Gauss--Bonnet cases  ($A$ and $GB$) are effectively two-derivative theories,
  and so  the associated
corrections can be read off directly from Eqs.~(\ref{Afe}) and (\ref{GBfe}).
We then have $\;\eta_A=\eta_E\left[1-40\alpha\right]\;$
and $\;\eta_{GB}= \eta_E\left[1-8\lambda\right]\;$; both in agreement
with the already-known results.~\footnote{For
$\eta_{GB}$, see ({\it e.g.}) \cite{BLMSY-0712.0805}. That
the scalar-squared case ($A$) is a match,
even if not documented explicitly,  follows from
the confirmation of $\;\eta_A/s_A= \eta_E/s_E\;$,
as must be true for any $f({\cal R})$ theory. Note that $\;s_A=s_E\left[1-40\alpha\right]\;$
follows from Wald's formula \cite{wald1,wald2}.}

However, for  the Ricci and Riemann-tensor squared cases
($B$ and $C$), there are now  $\Box^2$ terms
present in the field equations.
Following the standard prescription, one would conclude
from Eqs.~(\ref{Bfe}) and (\ref{Cfe}) that $\;\eta_B=\eta_E\left[1-8\beta\right]\;$
and $\;\eta_{C}= \eta_E\;$,  which
are indeed in perfect agreement with those obtained from
any of the previously described
methods.~\footnote{Similar to
case $A$, the Ricci-square result follows from knowledge of
the entropy density via Wald's formula and that $\;\eta_B/s_B= \eta_E/s_E\;$
\cite{BLMSY-0712.0805}.  Also, to avoid any confusion, the
ratio $\eta_C/s_C$ changes by an overall factor of
$\;1-8\gamma +{\cal O}[\gamma^2]\;$,
same as for Gauss-Bonnet. Only that, for the Riemann-squared case,
this correction comes entirely from the entropy \cite{BM-0808.3498}.}

On the horizon,  $\;\Box h_{xy}\;$ and the background field
equation $\;\Box_E h_{xy}\;$ are interchangeable up to order $l_p^2/L^2$.
Then, as any $\Box$  must be of order $l_p^2/L^2$,
$\;\Box^2 \ll \Box\;$ certainly follows. Off the horizon the situation changes, since it is $\;\Box_E h_{xy} =[\Box + 2 F(r)] h_{xy}\;$ that vanishes on shell, and so  $\Box^2\ll \Box\;$ is no longer valid. A single $\Box^2$ term now makes a generic contribution of
\begin{equation}
\label{offhor}
\;\Box^2=\Box^2_E -4 F \Box_E+\cdots\;,
\end{equation}
where the dots signify that we are disregarding mass terms as usual.
In terms of our generic theory of Eq.~(\ref{generic}),
$\;b\to b- 4 a F\;$. The $\Box^2$ term has, through
its prefactor  $a$, entered into the calculation of the kinetic coefficient.~\footnote{This
distinction between the horizon and other surfaces  seems contradictory
to the claim that $\eta_{FT}$ can be calculated
at any radius in the spacetime \cite{LI}. However, this claim has only been established
for two-derivative theories.}

It would seem that the horizon value of $b$ and, hence,
$\eta$ are insensitive to the $\Box^2$ terms.  We will, contrary to this expectation, show  that
the $\Box^2$ terms are always implicated with
ghosts; with these
directly influencing the value of $b$ throughout the spacetime ---  including
at the horizon!

\subsection{Ghosts in the machinery}

It  is well known that the presence of  a $\Box^2$ in the equation of motion signals that a ghost lurks in the theory. The usual argument for ignoring this is that the Planck-scale ghost becomes infinitely massive once the ultraviolet cutoff has been sent to infinity \cite{robbed}. However, for the case at hand, $\epsilon\sim l_p^2/L^2$ needs to be kept finite. If one takes the strict decoupling limit $\epsilon=0$ and all corrections originating from the higher-derivative terms vanish, as expected.

Let us  begin here by exposing the ghosts. We will do so using a generic four-derivative theory,  as described by Eq.~(\ref{generic}), to emphasize that the ghosts are neither model nor radius specific. First,   Eq.~(\ref{generic}) can be recast, using Eq.~(\ref{offhor}), into
\be
\left[\Box_E  \;+\;\epsilon a L^2\Box_E^2 \;
+\;\  \epsilon{\tilde b}\Box_E\right]h_{xy}\;=\;0\;,
\label{generic2}
\ee
where  ${\tilde b}\equiv b-4aF$
is a radially dependent parameter that generally differs from $b$  but
$a$ is the exact same number as before. Or, up to leading order in $\epsilon$,
\be
(1+ \epsilon b)\Box_E\left[1\;+\; \epsilon a L^2\Box_E \right]h_{xy}\;=\;0\;,
\label{dem3}
\ee
with the tilde on $b$ henceforth implied.

The factorization in Eq.~(\ref{dem3}) implies that, at order $\epsilon$, there are two separable modes in this problem. Let us then write the graviton as $\;H = H_0 + H_1\;$ (with tensor indices implied), where  these are meant  (and shortly will be shown) to satisfy
\be
\Box_E H_0 \;=\; 0\;,
\label{newby1}
\ee
\be
\left[\epsilon a \Box + \frac{1}{L^2}\right] H_1 \;=\; 0\;.
\label{newby2}
\ee

To explicitly expose the two distinct  modes,
let us invoke the standard practice in flat spacetime of inverting the linearized operator and then factorizing. That the inversion process extends in a straightforward manner from flat space to AdS space
follows from the fact that $\Box_E$ is  a scalar. Then $\;\Box=-p^2\;$ of flat space is replaced with an appropriately generalized momentum; say  $\;\Box_E=-{\widetilde p}^{\ 2}\;$.
On the horizon of a black brane, $\;{\widetilde p}^{\ 2}=p^2\;$. Then,
\be
\frac{1}{\Box_E\left[1\;+\; \epsilon a L^2 \Box_E\right]  }
\;\;=\;\;
\frac{1}{\Box_E}\;\;
-\;\; \frac{1}{\Box_E\;+\; (\epsilon a L^2)^{-1}}
\;
\label{dem4}
\ee
The first term on the right can be identified as the propagator of the massless Einstein graviton. Meanwhile, the second term
is the propagator of a massive ghost graviton, as evident from
the  ``wrong'' sign  relative to the former.
This outcome does not depend on  the sign nor magnitude  of
the perturbative coefficient  $\epsilon$,
which  only determines the mass of the ghost. Let us recall that $\;a,b\sim {\cal O}[1]\;$ and $\;\epsilon\sim l^2_p/L^2\;$, and so  the ghosts have a  squared mass $M^2 \sim (\epsilon L^2)^{-1}\sim 1/l_p^2$.
As already stressed at the end of Section 2, we are looking at corrections to leading order in $\epsilon$. Consequently, the very massive ghosts may indeed affect the result.

If a four-derivative  theory does have ghosts,  these persist throughout
the entire spacetime because of
the invariance of the dimensionless constant $a$ associated
with the $\Box^2$ term.~\footnote{The same claim can, in fact, be
made for any  higher-derivative theory that depends only the Riemann
tensor. The inclusion of other types of fields could jeopardize
the invariance of $a$, but  this parameter would still be typically
non-vanishing at all radii.}
Hence, the ghosts are not confined to the AdS  bulk interior
and  can contribute to the field-theory
viscosity to leading order in $\epsilon$.

Now suppose that we want to restore the normalization factor
of $(1+b)$
to the mode equations (\ref{newby1},\ref{newby2}).
According to any of  the standard methods,
the factor $(1+\epsilon b)$ is strictly attributed to the massless
graviton. This follows
from $\eta$ always going as the value of
$(1+\epsilon b)$ at the horizon (see Subsection~3.1) and the implicit assumption that
any ghost
mode would have ultimately decoupled.
On this basis,  the  mode equations should be reformulated
as
\be
(1+\epsilon b)\Box_E H_0 \;=\; 0\;,
\label{newby3}
\ee
\be
\left[\epsilon a \Box +\frac{1}{L^2}\right] H_1 \;=\; 0\;.
\label{newby4}
\ee
But, as $\;b=b(r)\;$ does not commute with $\Box_E$, this formulation
is problematic in that it could not be obtained
through a process of factorization ({\it cf}, Eq.~(\ref{dem3})).
Hence, whatever might be the actual coupling for the massless graviton, it can be generically different than the standard result of $(1+\epsilon b)$.
This difference can be attributed to the fact that the gravitational coupling $(1+\epsilon b)$ can only be attributed to the sum of the modes and not to any single one of them. This argument shows that the ghost mode can indeed impact the calculation of  $\eta$.

In summary, we have seen that the ghosts have, through their association with the coupling correction $b$, managed to infiltrate into the coefficient of the kinetic terms and, thus, into the calculation of $\eta$. This is true insofar as the squared mass in units of the AdS radius $\;L^2 M^2\sim \epsilon^{-1}\sim L^2/l_p^2\;$ is kept large but finite. In the next section, we will propose a procedure that guarantees the complete decoupling of the ghost mode and allows for an unambiguous calculation of $\eta/s$.

\section{New (and improved) prescription for calculating $\eta$}
\label{section:neweta}

We conclude that, to proceed, one should  ensure that the influence of the ghosts is completely eliminated before the calculation of $\eta$ is carried out. This could be accomplished through a carefully imposed choice of BC's, as we explain next.

\subsection{An exorcizing boundary condition}

Let us reconsider the generic field equation (\ref{dem3}).
For a generic
extension of Einstein's theory, one can expect two BC's
per each order of $\epsilon$.  This enables one to individually fix
each order of the solution to fulfill the
stated conditions \cite{ALEX1}. Here, we have
rearranged the solution so that it is separated  according to
the different degrees of freedom rather than perturbative order. Nevertheless,
as is evident from Eqs.~(\ref{newby1}) and (\ref{newby2}),
both of the modes (massless graviton and ghost) still satisfy a simple
quadratic
equation and, thus, both have a corresponding incoming and
outgoing solution.  And so, with  four BC's at
our disposal, we are using one of these to stipulate that
the ghost is an  incoming plane wave at the horizon and
another
to kill off the ghost by, for instance, imposing the Dirichlet BC at the
horizon.~\footnote{Generically, these hydrodynamic modes are
regular everywhere, except at the horizon where regularity must be  imposed by
hand (see, {\it e.g.}, \cite{Kovtun:2005ev}). Hence, fixing the wavefunction to vanish
at the horizon is a  sufficient condition for the mode to be
vanquished throughout
the spacetime.}
The remaining two BC's are then used to  ensure that the massless graviton
satisfies the usual pair of BC's: Dirichlet (or vanishing wavefunction) at the AdS  boundary and incoming plane wave  at the black brane horizon.

Our new BC's are compatible with the standard set of
incoming plane wave at the horizon  and Dirichlet on a
radial shell~\footnote{As stressed in \cite{Shear}, there is nothing special about
the outer $AdS$ boundary in this regard, even if the standard choice.
The Dirichlet condition can readily be imposed on any radial shell
exterior to the horizon. This includes at the horizon when taken
as a  suitable limit of the stretched horizon.}
for the {\em total} wavefunction. Let us recall that the total solution can be expressed as $\;H=H_0+H_1\;$, with $H_{0/1}$  respectively labeling the massless
graviton and massive ghost. We also recall the
%Then, the explicit factorization of Eq.~(\ref{dem4})
%yields (up to ${\cal O}[\epsilon]$) a
relevant pair of  quadratic equations,
\begin{equation}
\Box_E H_0 \;=\;0\;,
\label{prop0}
\end{equation}
\begin{equation}
\left[\epsilon a\Box_E + \frac{1}{L^2}\right] H_1 \;=0\;\;,
\label{prop1}
\end{equation}
with both admitting plane-wave solutions.

Adopting the standard incoming  plane-wave solution and using the dimensionless (inverted) radial coordinate $\;u=r_h^2/r^2\;$, we can express the total wavefunction of the mixed-index gravitons $\;\Phi_{0/1}\equiv g^{xx}H_{0/1}\;$ as the sum
$\;\Phi(u,\omega, q)=\; {\cal C}_0\Phi_0 \;+\; {\cal C}_1\Phi_1\;$ such that
\be
\Phi_0 \;=\; [f(u)]^{-i\omega/2}\; \Upsilon_0(u,\omega, q) \;,
\ee
\be
\Phi_1 \;=\; [f(u)]^{-i\omega/2}\; \Upsilon_1(u,\omega, q) \;.
\ee
Here,  $\; f(u)=1-u^2\;$ (essentially, $|g_{tt}|$),
$\omega$ and $q$  represent a dimensionless frequency and wavenumber
(as defined in, {\it e.g.}, \cite{Shear}), ${\cal C}_{0/1}$ are
normalization constants and $\Upsilon_{0/1}$ are model-dependent
functions of $\;u$,$\omega$,$q\;$. The latter functions can be uniquely
fixed,
up to normalization, by imposing regularity at the horizon (with regularity
assured elsewhere).

The hydrodynamic limit implies  $\;\omega$,$q\ll 1\;$,
so that --- in practice --- one expands $\Phi_{0/1}$ out in increasing powers
of $\omega$ and $q$  and considers only the first few terms in
the series.  For instance, Einstein gravity ($\;\epsilon=0\;$)
leads to
$\;\Upsilon_0 = 1+{\cal O}[\omega^2,q^2]\;$
\cite{Kovtun:2005ev}.~\footnote{At the horizon,
either of our modes has  precisely this same form. For the massless
graviton, this is trivially so at any radius. For the massive ghost,
this follows from
the red shift at the horizon making this mode effectively massless.}
Technically, however, the hydrodynamic limit should
only be applied at the very end of the calculation.

For $\Phi_0\;$,  the  Dirichlet BC is imposed at the outer boundary ($\;u\to 0\;$)
and then the dispersion relation $\;\omega=\omega(q)\;$ is fixed to attain
\be
\left.[f(u)]^{-i\omega/2}\;\Upsilon_0(u,\omega, q)\right|_{u\to 0}=0\;.
\ee
Meanwhile, the normalization  ${\cal C}_0$ is
fixed in the standard way
\be
{\cal C}_0^{-1} \;=\;
\left.[f({\overline u})]^{-i\omega/2} \;
\Upsilon_0({\overline u},\omega, q)\right|_{{\overline u}\to 0}\;,
\ee
where ${\overline u}$ is a constant ``cutoff'' scale.
Consequently,  the renormalized wavefunction
$\;{\widehat \Phi}_0\equiv {\cal C}_0 \Phi_0$ satisfies
\begin{equation}
 \left. {\widehat \Phi}_0 \right|_{u\to 0} \;=\; 1\;.
\end{equation}

For the ghost mode $\Phi_1$, we now deviate from the
usual convention and rather fix  its normalization
with the constraint $\;{\cal C}_1=0$. Since regularity has already been imposed
at the horizon through the correct choice of the functional form
$\Upsilon_1$, the  renormalized wavefunction
$\;{\widehat \Phi}_1\equiv {\cal C}_1 \Phi_1$ is compliant with
the horizon Dirichlet  BC
\begin{equation}
 \left. {\widehat H}_1 \right|_{u\to 1} \;=\; 0\;.
\end{equation}
As a result, the total normalized wavefuntion $\;{\widehat \Phi}=
{\widehat \Phi}_0 + {\widehat \Phi}_1\;$ reduces to ${\widehat \Phi}_0$,
\begin{equation}
{\widehat \Phi}(u,\omega,q)\;=\;  \left.
\left[\frac{f(u)}{f({\overline u})}\right]^{-i\omega/2}
\frac{\Upsilon_0(u,\omega, q)}{\Upsilon_0({\overline u},\omega, q)}
\right|_{{\overline u}\to 0}\;.
\end{equation}
In this way, we have ensured that the amplitude of the heavy ghost mode is made to vanish
without insisting on the  vanishing of $\epsilon$.

The effect of imposing the
BC's that ``exorcise'' the ghost is to
set the right-most term in Eq.~(\ref{dem4}) to vanish.
We can then re-invert the remaining operator
to obtain  a ``unitarized" version of
the field equation (\ref{dem3}) in which the rightmost factor has been set to unity:
\be
\left(1+\epsilon  c\right) \Box_E\;
h_{xy}\;=\;0 \;,
\label{dem5}
\ee
where the  order-unity radial function $c(r)$ is related to but
not necessarily equal to the original radial function $b(r)$.
As already discussed, this normalization factor
may well  have  changed in the described process.

Because the value of the function $c$ could have been ``infected'' by the ghost mode, it is desirable to bypass the individual evaluation of $\eta$ and $s$,
which  would require an exact calculation of $c$. Rather, we will calculate the ratio $\eta/s$ from which the exact value of the normalization function $c$ actually drops.

The field equation Eq.~(\ref{dem5}) can be derived from an effective action with the following Lagrangian:
\be
{\cal L}_{eff}\;=\;-\frac{1}{2}
\left(1+\epsilon  c\right){\overline g}^{pq}{\overline \nabla}_p h^{xy}\;
{\overline \nabla}_q h_{xy}+\cdots\;.
\label{above}
\ee
Since this is now a two-derivative theory of gravity,
we can directly invoke  Eq.~(\ref{cai-eqn}) to determine
the shear viscosity; hence, given $c$, $\eta$ goes as
$\;(1+\epsilon c)\eta_E\;$.

Now, what about  the entropy density?  Once the ghosts
have been exorcized, the gravitational theory is Einstein's with a renormalized gravitational coupling. This follows from \cite{DH1,DH2,DH3}, where it has been made clear
that a consistent two-derivative theory for a massless spin-two
field is uniquely described by Einstein gravity.
In general,
differently polarized gravitons would ``perceive'' different
values for the couplings. However, for a consistent two-derivative theory, there
can be no such distinction between the polarizations \cite{DH1,DH2,DH3}.
So that, by eliminating the ghosts,
we have also rendered  gravity to be  polarization indifferent.

We can then call upon the analysis from Section~III
of  \cite{BGH-0712.3206}, where the Wald
entropy has been reinterpreted as an effective coupling for
the $r,t$-polarized gravitons,
to  deduce that  $s$ goes as
$\;(1+\epsilon c)s_E\;$.
Meaning that the ratio $\eta/s$ must necessarily be
equal to the Einstein value of $1/4\pi$. That is, the ``ghost-busted'' theory may  have a shifted value of $\eta$,  but it comes  with the assurance that $s$ will be corrected from its Einstein value   by the very same amount.

\subsection{Comparison to another choice of boundary conditions}

We would like to  be precise on how our newly proposed prescription
differs from those previously used; in particular, the effective-action
method \cite{DUDE} and those similar to it
({\it e.g.}) \cite{ROB}. Given a four-derivative theory,
a prescription
for calculating $\eta$
is a three-step process: (i) find the equations of motion, (ii)
solve the equations of motion, (iii) decide on how to extract the viscosity.
There obviously  can be no dispute on the first step and,
given that the action has been consistently
reduced to
a two-derivative form, there should be no contention about  the third.
Where our new prescription then differs is only
in the second step,
where a clear distinction has been made between the appropriate choice of  BC's and, hence, on the
resulting solution. It is interesting to compare these two sets of
solutions in detail, as we do next.

Let us begin here by recalling
Eqs.~(\ref{dem5}) and (\ref{prop1}); these being the field equations
 for the (post-normalized)  massless graviton
 $H_0$ and the ghost graviton $H_1$, respectively:
%$H_1$ respectively, we essentially have the following two equations:
%\begin{equation}
%\left[1 +\epsilon b \right]\Box_E H_0 \;=\;0\;,
%\end{equation}
%\begin{equation}
%\left[\epsilon a\Box_E + \frac{1}{L^2}\right] H_1 \;=0\;\;.
%\end{equation}
%Notice that the leading-order form of these equations,
\begin{equation}
(1+c)\Box_E H_0 \;=\; 0\;,
\label{stuff}
\end{equation}
\begin{equation}
\left[\epsilon a\Box_E + \frac{1}{L^2}\right] H_1 \;=0\;\;.
\end{equation}
%does not depend
%on the choice of factorization, as per the discussion
%at the end of  Subsection~3.3.

We want to compare these equations
%Eqs.~(\ref{dem5}) and (\ref{prop1})
with those obtained in the ``usual''
manner by which one expands the solution according to perturbative order
(rather than  isolating the degrees of freedom).
Let us label the zeroth-order and first-order modes respectively as $h_0$ and
$\epsilon h_1$. Then, the previous field equations translate into
\begin{equation}
\Box_E h_0 \;=\; 0\;,
\end{equation}
\begin{equation}
 \left[ \epsilon a L^2 \Box_E + \epsilon b\right]\Box_E  h_0 \;+\; \epsilon \Box_E h_1 \;=\; 0\;.
\end{equation}
The standard choice of BC's \cite{ALEX1} is incoming
at the horizon and Dirichlet  at the outer boundary for  both $h_0$ and $h_1$.

To avoid clutter, let us consider the simple but instructive  case of
$\;b=c\;$.
Comparing the two sets of field equations, one finds that compatibility
(to first-perturbative order) requires the following dictionary:
 \begin{equation}
h_0 \;=\; H_0 + \frac{L^{-2}}{\Box_E}  H_1 \;,
\end{equation}
\begin{equation}
h_1 \;=\; - a L^2 \Box_E H_0 - \frac{L^{-2}}{\Box_E} b H_1 \;,
\end{equation}
so that the two sets of modes share a non-local relationship.
%for any choice of  $\; b_{0} +b_{i} =b\;$ and
%irrespective of how one chooses the
%initial factorization.
%Another choice of initial factorization would have only  changed
%the relative  contributions that $H_0$ and $H_1$  make
%to the higher-order mode $h_1$.
%This recurring freedom in the separation of  $b$
%again makes it clear that the ghost mode is implicated in
%the coefficient of the kinetic term.
Now, it follows from the above equations and $\;\Box_E\sim\epsilon\;$
 (so $\;\frac{1}{\Box_E}\sim \frac{1}{\epsilon}\;$) that
$H_0$ is $h_0$ with a perturbative correction whereas $H_1$ is
strictly is of order $\epsilon$, as it must be since the ghost is required
to vanish in the Einstein or  $\;\epsilon\to 0\;$  limit.
% \begin{equation}
%H_0 \;=\; h_0 + b h_1 \;,
%\end{equation}
%\begin{equation}
%H_1 \;=\; a L^4 \Box_E^2 h_1 \;.
%\end{equation}
%Hence,  $H_0$ is $h_0$ with a perturbative correction and $H_1$ is
%strictly is of order $\epsilon$, as it must be since the ghost is required
%to vanish in the Einstein or  $\;\epsilon\to 0\;$  limit.
%Notice that $h_0$ and $h_1$ are both of order unity in this set-up.
%This too is consistent, since $h_1$ already contributes at
%${\cal O}[\epsilon]$
%in the field equations.

%One further point is that the
%two sets of modes  share a non-local relationship;
%however, one can, if so desired, use the zeroth-order field equation
%to convert these relationships  into an explicitly local form. In this way, $H_1$ is
%really just $h_1$ up to some  radially dependent factor.

Comparing our exorcizing BC's to the standard set,  we see that the difference between them is quite subtle, as it only appears  at order $\epsilon$ for both $h_0$ and $h_1$.
From the discussion in the preceding subsections, it is clear that the values of $\eta$ and of $\eta/s$ are sensitive to the choice of BC's. In particular, an order-$\epsilon$  difference  in the BC's  for the massless graviton results in an equally small but still significant change in the value of $\eta/s$.

\subsection{General arguments}

For theories leading to equations of motion with more than four derivatives, it is quite possible that not all of the additional degrees of freedom are ghosts (see the discussion on six-derivative equations at the end of this section);
meaning that the ghost-reduced action need not be  limited to two derivatives nor even be local. The latter because our prescription would, in general, yield a non-local
field equation due to summing over two or more inverted operators.

For such cases, we propose proceeding as follows:  First,
after re-inverting the sum of the non-ghost propagators,
one should convert this ghost-free field equation into a local form by  expanding the propagators of massive modes (with masses $M_i$) in terms of $\;\Box/M_i^2 \ll 1\;$. Then, following the
effective-action method \cite{DUDE}, one should reduce the order of this field equation by treating the higher-derivative terms perturbatively while
iteratively applying a lower-order  field equation and
the BC's.  However,  the initial field equation for starting the  iterative
purposes is, in general, no longer that of the background.
It is, rather, the equation one  obtains by
perturbatively expanding and then suitably truncating the
ghost-free field equation. Finally, one should, as before,  translate
the reduced field equation into a two-derivative effective action
and then apply Eq.~(\ref{cai-eqn}).

One might then wonder as to how  $\eta/s$
would  turn out for such  higher-derivative theories. We will argue below
that, on general grounds, this ratio can now change but only in such a
way that it increases relative to its Einstein value

We have advocated  elsewhere \cite{us-new} that any gravitational coupling,
such as $\eta$,  can only increase from its Einstein value for a unitary extended theory.
This is because any consistent extension will necessarily introduce new degrees of freedom to supplement the Einstein graviton. Given that these
are unitary, the extended theory can only introduce additional channels that act to  increase the couplings.

Let us recall the relevant discussion from \cite{us-new}.
Given our gauge choice,   the $h_{xy}$ gravitons
can only couple linearly  to other particles of spin 2 {\it and}  of the same polarization.
Near the horizon, the associated
1PI graviton propagator~\footnote{See ({\it e.g.}) \cite{Dvali3} for the   fully decomposed form of the graviton propagator. Note that this flat-space form
of the propagator is appropriate near any horizon --- including that of
a brane in AdS space ---
as any horizon  has an effectively flat geometry.  This is
so because $\;\Box_E\propto-\partial^2_t +\partial^2_{r_{\ast}}\;$  as $\;r\to r_h\;$  (where $\;r_{\ast}\equiv\int dr/F\;$ is  a generalized ``Tortoise'' coordinate).}
takes on a particularly simple form:
\be
\langle h_x^{\;y}(q)h_y^{\;x}(-q)\rangle \;=\;\frac{\rho_E(q^2)}{q^2}\;+\;\sum_{i}\frac{\rho_i(q^2)}{q^2+m_i^2}
\;,
\label{help}
\ee
where $\;q^2=-q^{\mu}q_{\mu}\;$ is the spacelike momentum,  $\rho_E$  is the gravitational
coupling for  the Einstein graviton,  and the $\rho_i$'s   are the couplings for the additional  spin-2 particles of
an extended theory --- any of which could be massive ($\;m_i^2\neq 0\;$) or massless
($\;m_i^2=0\;$).~\footnote{By the same reasoning as in the prior footnote,
any particle near the horizon is effectively massless and ``perceives''
an effective $2D$ geometry. Hence, one can set all  the $m_i^2$ to zero and need only consider the
$t$ and $r_{\ast}$ components of $q^{\mu}$. Note, though, that the full $5D$ tensorial
structure of the gravitons is still maintained. See
\cite{Brustein:2010jb} for further explanation.}
The couplings can, as indicated, depend on the energy scale $q\;$; and
we work in units such that $\;\rho_E(0)=1\;$  correctly fixes the Newtonian force
at large distances.

The essential point here is that the couplings are really spectral densities
of the schematic form $\;\rho=\sum_n \langle 0|h|n \rangle\langle n|h|0 \rangle\;$ \cite{sred} and,  as such, assured
to be positive as long as all the inserted states have a positive norm.
Meanwhile, the Einstein  coupling
can itself  be modified in either direction ({\it cf}, Eq.~(\ref{dem5})), but this is an illusionary effect,
as one would always recalibrate her instruments  to maintain  the $\;\rho_E(0)=1\;$
normalization.   In any event,  any such  modification to
the Einstein coupling
would immediately
cancel out of the  ratio $\eta/s$, as Einstein gravity must be insensitive
to the polarization of  gravitons (which is just a restating of the equivalence principle).
Conversely, the ``non-Einstein''  gravitons are generally sensitive
to the polarization but can still only act to increase
$\eta/s$  from the Einstein value of $1/4\pi$.

To sum up, the microscopic theory is, on the basis of unitarity,
telling us that $\eta/s$ can only increase from the Einstein value.
This point allows us to clarify what was previously meant by
``considerations of unitarity''.
For addressing such matters of  principle,
a Planck-scale ghost is obviously  unreliable,
as would be the case for any  mode
near or above the effective-theory cutoff. On the other hand,
including  the ghost in the calculation enables $\eta/s$ to
change freely in any direction.
Hence, we conclude that, to obtain results from the effective theory that
are consistent with microscopic unitarity, it is necessary to impose
BC's that eliminate these unreliable modes.  Consistency then
requires that
a physically meaningful quantity such as
$\eta/s$ should no longer be sensitive to cutoff-scale physics.
%this ratio should not depend on the coefficient $a$
%of the $\Box^2$ term in a four-derivative theory.

Applying the preceding ideas  to our generic four-derivative model,
we expect   that the ratio $\eta/s$  must
remain independent of   the perturbative coefficient  $\epsilon$  at linear order. If this were not so, one  could always reverse the direction of  the $\eta/s$  correction by simply changing the sign of $\epsilon$. Reassuringly, this is exactly what our revised prescription ensures!  After all, the non-Einstein degree of freedom  is inevitably
a ghost ({\it cf}, Eq.~(\ref{dem4})); so  that, with our  choice of
BC's, the physically relevant theory
is guaranteed to reduce to Einstein's.

\subsection{Six and higher-derivative equations of motion}

Let us now discuss how the situation that we have described in such detail for the case of four-derivative theories is modified for theories leading to six-derivative equations of motion.
We will only provide a sketch, relegating a detailed discussion to a subsequent paper.  Six or higher-derivative equations can be quite different from four-derivative theories.
With six derivatives, for example, one would obtain two additional degrees of freedom,
only one of which need be a ghost.  To illustrate this, let us consider   ``toy-model''  versions
of  a six-derivative linearized operator. For instance,
$\;{\cal G}_1\equiv\Box_E
+\zeta\Box_E^2\left(c-\frac{1}{c}\right)-\zeta^2
\Box_E^3 \;$ (with $\zeta$ a perturbative coefficient,
$c\neq 1$ parameterizing  mass and units of $L=1$),
which factorizes to give
\be
\frac{1}{{\cal G}_1}\;\;=\;\;\frac{1}{\Box_E} \;\;-
\;\;\frac{c^2}{c^2+1}\frac{1}{\left[\Box_E+\frac{1}{c\zeta}\right]}
\;\;-\;\;\frac{1}{c^2+1}\frac{1}{\left[\Box_E-\frac{c}{\zeta}\right]}\;.
\ee
In this case, both of the massive gravitons are ghosts and should be
eradicated from the calculation. On the other hand,
$\;{\cal G}_2\equiv\Box_E
-\zeta\Box_E^2\left(c+\frac{1}{c}\right)+\zeta^2\Box_E^3\;$
leads to
\be
\frac{1}{{\cal G}_2}\;\;=\;\;\frac{1}{\Box_E} \;\;-
\;\;\frac{c^2}{c^2-1}\frac{1}{\left[\Box_E-\frac{1}{c\zeta}\right]}
\;\;+\;\;\frac{1}{c^2-1}\frac{1}{\left[\Box_E-\frac{c}{\zeta}\right]}\;.
\ee

This second case is more interesting because only one of the massive
gravitons is a ghost, with the choice depending on whether $c$ is greater
or less than unity.  For such a model, it can then be expected that $\eta/s$ does change,
as a new unitary degree of freedom would generally be
sensitive to graviton polarization.
Nonetheless,  seeing that  ${\cal G}_2$  only differs  subtly from  ${\cal G}_1$,
one is now able to envision a scenario whereby  the emergence of a new degree of
freedom is correlated with a strictly positive correction to
  $\eta/s$. Provided that the ghosts have been properly handled, it is our contention that this is, indeed, what must happen.~\footnote{One might be concerned
that $s$ --- also being  a type of gravitational coupling ---  should likewise
increase, possibly faster than $\eta$. Nonetheless, we have shown elsewhere
\cite{us-new}
that it is always possible to choose fields and coordinates such that $s$ is
calibrated to its Einstein value. Then, since $\;\eta\geq \eta_E\;$
is a  ``gauge''-invariant statement, the previous claim follows.}

\section{Gauss--Bonnet redux}
\label{GBR}

There is still an important example to be dealt with; the  Gauss--Bonnet
gravity model $\;{\cal L}_E+{\cal L}_{GB}\;$. This theory leads to a second, rather than fourth, order equation for the graviton as can be seen from Eq.~(\ref{GBfe}). Hence, the issue of the existence of ghosts or choice of boundary conditions becomes irrelevant. Obviously, unitarity considerations cannot be applied in a useful way in this case.

Yet,  we have found (as did past studies such as
\cite{BLMSY-0712.0805}) that
\be
\frac{\eta_{GB}}{s_{GB}}=\frac{1}{4\pi}\left[1-8\lambda\right]\;,
\label{GB}
\ee
as $s_{GB}$ is known to be functionally equivalent to $s_E$.
Evidently, the ratio will decrease below $1/4\pi$ whenever $\lambda$ is positive. So, this theory seems to provide a  counter-example to the main theme of our paper, as it apparently demonstrates a ghost-free
theory for which the viscosity--entropy ratio can be less than its Einstein value. However, this conflict is only an apparent one, as  we now explain.

There are two distinct ways of viewing the Lagrangian $\;{\cal L}_E+{\cal L}_{GB}\;$; as representing a theory unto itself  ({\it i.e.}, distinct from Einstein's) or as a perturbatively corrected version  of the Einstein Lagrangian. If one chooses the former point of view, then the microscopic theory is Gauss--Bonnet gravity and there can be no contradiction with what we have already said. All methods of calculations agree as they must, and the issue of imposing boundary conditions in a higher-derivative theory never arises.

 On the other hand, from the effective field-theory perspective,
Einstein's theory is viewed as the infrared limit of  the quantum gravity
theory and the various corrections  (${\cal L}_X$)
are  regarded as a consequence of integrating out heavy degrees of freedom.
Then, even if the Lagrangian has the finely tuned  Gauss--Bonnet form
($\;\beta=-4\alpha=-4\gamma\;$) at some scale, it will be of the generic form
at other scales and lead
to quartic equations of motion for the graviton.

So the issue now becomes an order of limits: Should one
first impose the Gauss--Bonnet fine-tuning condition and then calculate
$\eta/s$, leading to Eq.~(\ref{GB}) and possibly to values of $\eta/s$ lower than $1/4\pi$?   Or should one first eliminate the ghosts from ${\cal L}_B$, ${\cal L}_C$, then impose the Gauss--Bonnet fine-tuning condition  and only afterwards evaluate the ratio? If one has a reason to regard the Gauss--Bonnet theory  as fundamental, then the former choice is correct. With this interpretation,
Gauss--Bonnet gravity can no longer be regarded as a perturbative correction of Einstein's theory and, so,
is outside of the scope of our previous claims.
On the other hand, if the Gauss--Bonnet model is to be regarded as the leading-order modification of the Einstein
Lagrangian,  the latter choice should be made and the correct calculation
leads to  a renormalized Einstein theory with the outcome
 $\;\eta_{GB}/s_{GB}=1/4\pi\;$.

\section{Concluding discussion}
\label{conclusions}

Let us summarize: We have shown that, for higher-derivative
theories that extend Einstein gravity, earlier choices of BC's for the purpose of calculating  the shear viscosity have resulted in  contributions from ghosts
to the value of $\eta$.  We have proposed a different choice of BC's that
decouples the ghosts. An immediate consequence
of our proposal is that, when the extensions are limited to four derivatives,
the ratio of  shear viscosity to entropy density
saturates its Einstein value to leading order in the strength of the corrections; that is, saturates the KSS bound.
We have gone on to argue,  on general grounds related to unitarity,
that  our prescription  can be expected to enforce the KSS bound for theories leading to six and higher-derivative equations of motion and to all orders in the strength of the corrections.
In such general cases, the saturation of the  bound should  no longer  persist, as
any gravitational coupling would naturally increase (from Einstein's) after
additional unitary degrees of freedom have been introduced into the gravitational sector.
The validity of our claims under these more generic circumstances remains an interesting, open question, which we hope to address with explicit calculations.

\section{Acknowledgments}

We thank Alex Buchel, Misha Lublinsky, Yaron Oz, Rob Myers and Aninda Sinha for useful discussions. We thank Dan Gorbonos for alerting us to a sign error in the equations of motion~(\ref{wald-eq}) in a previous version of the paper and Merav Hadad for discussions on this issue.

The research of RB was supported by The
Israel Science Foundation grants no 470/06 and 239/10.
The research of AJMM was supported by the University of Seoul,
the Korea Institute for Advanced Study  and Rhodes University.
AJMM thanks  Ben-Gurion University for their hospitality during his visit.

\appendix

\section{Previous proposals}

Let us be more specific on how the shear viscosity has  previously been evaluated
for a higher-derivative theory. We begin with the ``effective-action'' method \cite{DUDE} and remind the reader that other approaches --- in particular, the generalized canonical-momentum treatment of \cite{ROB} ---  are know to yield identical results.

Let us consider the Lagrangian $\;{\cal L}={\cal L}_E+\epsilon{\cal L}_X\;$, where ${\cal L}_X$
is an arbitrary four-derivative extension and $\epsilon$ controls the
perturbation. To quadratic order in
the $xy$ gravitons of the metric, ${\cal L}$ can be written  as
\be
\sqrt{-g}{\cal L}\;=\;
{\widetilde A}(u)\phi^{\prime\prime}\phi
\;+\;{\widetilde B}(u)\phi^{\prime}\phi^{\prime}
\;+\;{\widetilde E}(u)\phi^{\prime\prime}\phi^{\prime\prime}
\;+\;{\widetilde F}(u)\phi^{\prime\prime}\phi^{\prime}
\;+\;\cdots \;,
\label{pm1}
\ee
where terms of lower-derivative order, as well as  terms vanishing in the hydrodynamic limit
of vanishing frequency and transverse momentum, are  denoted by ``$\cdots$''.
In this appendix, we use the radial coordinate $\;u=1/r^2\;$ and set $\;L=r_h=1\;$ so that the horizon is at $\;u_h=1\;$.
Additionally,
$\;\phi = h^y_{\ x}\;$,
a prime denotes a differentiation with respect
to $u$ and
${\widetilde A}$,${\widetilde B}$,${\widetilde E}$,${\widetilde F}$ indicate model-dependent radial functions.
(For Einstein gravity, $\;{\widetilde E}={\widetilde F}=0\;$.)
The background metric (\ref{metric}) becomes
\be
ds^2\;=\;-\frac {f(u)}{u}dt^2\;+\; \frac {du^2}{4u^2 f(u)}
\;+\;\frac{1}{u}\left[dx^2\;+\;dy^2\;+\;dz^2\right]\;,
\label{pm1a}
\ee
with $\;f(u)=1-u^2\;$ and $\;f(1)= 0\;$.

As a first step, one integrates by parts (with the relevant surface terms presumed to exist \cite{ALEX}) to obtain
\be
\sqrt{-g}{\cal L}\;=\;
{\widetilde E}\phi^{\prime\prime}\phi^{\prime\prime}
\;+\;\left[ {\widetilde B}-{\widetilde A}-\frac{{\widetilde F}^{\prime}}{2}\right]\phi^{\prime}\phi^{\prime}
\;+\;\cdots \;.
\label{pm2}
\ee
The associated field equation is then
\be
\left[{\widetilde E}\phi^{\prime\prime}\right]^{\prime\prime}\;+\;\left[{\widetilde A}
-{\widetilde B}+ \frac{{\widetilde F}^{\prime}}{2} \right]\phi^{\prime\prime}
\;+\;\left[ {\widetilde A}-{\widetilde B}+ \frac{{\widetilde F}^{\prime}}{2} \right]^{\prime}\phi^{\prime}\;+\;\cdots \;=\;0\;.
\label{pm3}
\ee

Next,
one is instructed to use the zeroth-order or Einstein field equation,
$\;\phi^{\prime\prime}=-Z^{\prime}\phi^{\prime}+\cdots\;$
with $\;Z\equiv\left[\ln(\sqrt{-{\overline g}}{\overline g}^{uu})\right]^{\prime}\;$, to reduce the higher-derivative ${\widetilde E}$ term, as  ${\widetilde E}$
is already of order $\epsilon$.
Twice applying the zeroth-order  equation  in the prescribed manner, one
ends up with
\be
\left(\left[(Z{\widetilde E})^{\prime}-Z^2{\widetilde E}+{\widetilde B}-{\widetilde A}-
\frac{{\widetilde F}^{\prime}}{2}\right]
\phi^{\prime}\right)^{\prime}\;+\;\cdots
\;=\; 0\;,
\label{pm4}
\ee
and the omitted terms now
also include those of order $\epsilon^2$.

Eq.~(\ref{pm4}) implies  an effective  Lagrangian of the following form:
\be
\sqrt{-g}{\cal L}_{eff}\;=\; -\frac{1}{{\overline g}^{uu}}
\left[(Z{\widetilde E})^{\prime}-Z^2{\widetilde E}+{\widetilde B}-{\widetilde A}-
\frac{{\widetilde F}^{\prime}}{2}\right]{\overline g}^{uu}
\partial_u\phi\partial_u\phi
\;+\;\cdots\;.
\label{pm5}
\ee
The shear viscosity can be identified with the
coefficient of the  kinetic term for the   $\phi$'s at the horizon
\cite{BM-0808.3498,LI,CAI}, and so
one would deduce that
\be
\eta\;\propto\; -\frac{1}{{\overline g}^{uu}}\left[(Z{\widetilde E})^
{\prime}-Z^2{\widetilde E}+{\widetilde B}-{\widetilde A}-
\frac{{\widetilde F}^{\prime}}{2}\right]_{|_{u\to 1}}\;,
\label{pm6}
\ee
where the horizon limit is imposed only at the end.

Let us confirm that this result is indeed equivalent
to that obtained from the canonical-momentum formalism of  Myers {\it et. al.} \cite{ROB}. Generalizing the methodology  of \cite{LI},
these authors have essentially identified
$\eta$ with  the (inverse) ratio of  $\phi$ to its canonical conjugate, as
computed at the AdS boundary and  in the hydrodynamic limit. A nice consequence of the latter limit
is that the same identification can be made  on  any radial surface in
the  spacetime,
including at the black brane horizon.
In this manner, for a generic four-derivative theory,
the shear viscosity has been formulated
strictly in terms of the horizon values of
${\widetilde A}$,${\widetilde B}$,${\widetilde E}$,${\widetilde F}$ and the metric.  Retrieving their formula from Eq.~(3.22) of  \cite{ROB}, we obtain (after some minor manipulations)
\be
\eta\;\propto\; -\frac{1}{{\overline g}^{uu}}\left[Y\left(\frac{{\widetilde E}}{Y^2}Y^{\prime}\right)^{\prime}
+{\widetilde B}-{\widetilde A}-\frac{{\widetilde F}^{\prime}}{2}\right]_{|_{u\to 1}}\;,
\label{pm6a}
\ee
where $\;Y\equiv \sqrt{-{\overline g}_{tt}{\overline g}^{uu}}\;$.  Comparing the last two equations, one can see
that it is enough to establish the equivalence of
$\;{\cal E}_1\equiv (Z {\widetilde E})^{\prime}-Z^2{\widetilde E}\;$
and $\;{\cal E}_2\equiv Y\left[\frac{{\widetilde E}}{Y^2}Y^{\prime}\right]^{\prime}\;$ on the horizon.
For this purpose, it is useful to recall that ${\widetilde E}$ is the coefficient of the four-derivative
term ({\it cf}, Eq.~(\ref{pm1})); meaning that it generically contains two  factors
of $\;{\overline g}^{uu}\sim  Y\;$ and is quadratically vanishing on the horizon.  Hence, we can rather work with
$\;W \equiv {\widetilde E}/f^2\;$ ($\;f\sim Y\;$),  with $W$   assured  to be regular as $\;u\to1\;$.

Let us now specialize to the background metric of Eq.~(\ref{pm1a}),
although, as one can check, this is not a necessary requirement.
Making the substitutions $\;{\widetilde E}=f^2 W\;$, $\;Z=\frac{f^{\prime}}{f}-\frac{1}{u}\;$,
$\;Y=2\sqrt{u}f\;$ into ${\cal E}_{1}$ and ${\cal E}_2$, one finds after simplifying that
\be
\frac{1}{f}{\cal E}_1\;=\; f^{\prime\prime}W\;+\;\left(f^{\prime}-\frac{f}{u}\right)W^{\prime}\;,
\label{pm6b}
\ee
\be
\frac{1}{f}{\cal E}_2\;=\; \left(f^{\prime\prime} -\frac{3}{4}\frac{f}{u^2}\right)W
\;+\; \left(f^{\prime}+\frac{1}{2}\frac{f}{u}\right)W^{\prime}\;.
\label{pm6c}
\ee
Finally, after imposing the horizon limit, one ends up with
\be
\frac{1}{f}{\cal E}_{1,2} \;\to\; f^{\prime\prime}(1)W(1) \;+\; f^{\prime}(1)W^{\prime}(1)\;
\label{pm6d}
\ee
in either case.

It is instructive to focus on a specific gravity theory. Let us consider,
for instance, the Riemann (tensor) squared  model or
$\;{\cal L}={\cal L}_E+{\cal L}_C\;=\;{\cal R}+12
+\gamma{\cal R}_{abcd}{\cal R}^{abcd}\;$, for which the background
metric (\ref{pm1a}) is
appropriate and
the explicit forms of ${\widetilde A}$,${\widetilde B}$,${\widetilde E}$,${\widetilde F}$ are already given in
Eq.~(3.22) of \cite{ROB}.~\footnote{Their $c_1$ is the same as our $\gamma$ and
they denote ${\widetilde A}$,${\widetilde B}$,${\widetilde E}$,${\widetilde F}$ without tildes.
Otherwise, the conventions are basically in agreement.}
Substituting these into Eq.~(\ref{pm6}) or Eq.~(\ref{pm6a})
and then simplifying, we have
\be
\eta \;\propto \; 1\;+\;8\gamma \left[f^{\prime}(1)-f^{\prime\prime}(1)\right]\;,
\label{pm7}
\ee
where the Einstein term has been ``normalized'' to unity and note that
the correction term is, for this particular background, a vanishing quantity.

Let us next consider the approach where one
continues to identify $\eta$ with the horizon coefficient of the kinetic
term for the $xy$-polarized gravitons in  the original action.
The premise being that the  gravitational coupling for a given class of gravitons  can be extracted from
their  kinetic terms \cite{BGH-0712.3206} and that the shear viscosity
is a measure of  the coupling for the $xy$ gravitons.
One can, in principle, make this identification by expanding out
the Lagrangian to quadratic order in $h_{xy}$.
It is, however, simpler to extract the kinetic term by a method proposed in  \cite{BM-0808.3498}, which can be viewed
as a   generalization of  Wald's entropy formula
\cite{wald1,wald2,BGH-0712.3206}
and is tantamount
to varying the Lagrangian by $\;{\cal R}^{xy}_{\;\;\;xy}\;$.
For the just-discussed Riemann-squared model,
this implies
\be
\eta\; \propto \; 1 \;+\; 4\gamma \left[{\overline {\cal R}}^{xy}_{\;\;\;xy}\right]_{|_{u\to 1}} \;=\; 1\;-\; 4\gamma f(1)\;.
\label{pm8}
\ee
Here, the correction term is also vanishing but the functional form apparently differs
from that of the previous computations. Nevertheless, we can help to
establish the claimed
equivalence (for four-derivative theories) as follows:

First, let us  expand the curvature component out as
$\;{\overline {\cal R}}^{xy}_{\;\;\;xy}={\overline {\cal R}}^x_{\ x}-{\overline {\cal R}}^{xu}_{\;\;\;xu}-{\overline {\cal R}}^{xt}_{\;\;\;xt}
-{\overline {\cal R}}^{xz}_{\;\;\;xz}\;$. Next, we use the equality $\;{\overline {\cal R}}^{x}_{\ x}={\overline {\cal R}}^{t}_{\ t}\;$,
which follows from the  spacetime being static  and  Poincare invariance on the brane.
Making this substitution and then expanding out ${\overline {\cal R}}^{t}_{\ t}$ in terms of
its four-index constituents, we have
$\;{\overline {\cal R}}^{xy}_{\;\;\;xy}={\overline {\cal R}}^{tu}_{\;\;\;tu}+{\overline {\cal R}}^{ty}_{\;\;\;ty}+{\overline {\cal R}}^{tz}_{\;\;\;tz}
-{\overline {\cal R}}^{xu}_{\;\;\;xu}-{\overline {\cal R}}^{xz}_{\;\;\;xz}\;$. Straightforward evaluation
at the horizon  then
leads to $\;{\overline {\cal R}}^{xy}_{\;\;\;xy}=2\left[f^{\prime}(1)- f^{\prime\prime}(1)\right]\;$.
Inserting this into Eq.~(\ref{pm8}), we verify a functional form for $\eta$ that agrees perfectly
with Eq.~(\ref{pm7}).

Such agreement
does not generally persist for theories with six or more derivatives.
For instance, let us consider the six-derivative corrected theory
$\;{\cal L}={\cal L}_E
+ \zeta  {\cal  R}_{abcd}{\cal R}^{cdbe} {\cal R}^{a}_{\ e}\;$,
for which  the effective-action method has already  been used in \cite{DUDE2}
to obtain
$\;\eta=\eta_E\left[1- 32\zeta\right]\;$.
On the other hand, extracting
the shear viscosity from the real action as prescribed
in \cite{BM-0808.3498}, one rather finds that
$\;\eta=\eta_E\left[1- 16\zeta\right]\;$.
Note that the equations of motion for this model are limited
to terms with no more than four derivatives, so that the two methods can
already be in  disagreement for theories leading to four-derivative
equations of motion.

%JOEY: I DO NOT UNDERSTAND THE FOLLOWING PARA.\\

%One might still wonder  as to
%how discarding the ghosts beforehand could change anything,
%seeing that Gauss--Bonnet gravity has zero net $\Box^2$ terms to begin with.
%Let us address this rather subtle point:
%For a ``conventional'' two-derivative theory, the variational
%problem is well defined,  so that the kinetic
%coefficient for the action or field equations
%has an unambiguous meaning.
%The same is, however, not true for a higher-derivative
%theory, for which the variational problem is either ill defined
%or non-existent.
%Lacking explicit knowledge about the boundary terms,
%one is left with a quandary
%when asking questions like ``what is the coefficient
%of the kinetic term?''.

\end{document}